\documentclass[conference]{IEEEtran}
\IEEEoverridecommandlockouts

\usepackage{amsmath,amssymb,amsfonts}
\usepackage{silence}
\usepackage{subcaption}
\usepackage{algorithmic}
\usepackage{algorithm}
\usepackage{array}
\usepackage{booktabs}
\newcommand{\tabitem}{~~\llap{\textbullet}~~}
\usepackage{tabularray}

\usepackage{rotating}
\usepackage{caption}
\usepackage{textcomp}
\usepackage{stfloats}
\usepackage{url}
\usepackage{nomencl}
\makenomenclature
\usepackage{pgfplots}
\usepackage{pgf-pie}
\usepackage{xcolor}
\usepackage{verbatim}
\usepackage{silence}
\usepackage[justification=centering]{caption}
\usepackage{graphicx}
\usepackage{cite}
\usepackage{hyperref}

\pgfplotsset{compat=1.18}

\def\BibTeX{{\rm B\kern-.05em{\sc i\kern-.025em b}\kern-.08em
    T\kern-.1667em\lower.7ex\hbox{E}\kern-.125emX}}

\begin{document}
	
	\title{Development of a New Type of Vortex Bladeless Wind Turbine for Urban Energy Systems\\
	}
	
\author{\IEEEauthorblockN{1\textsuperscript{st} Dongkun Han{*}}
	\IEEEauthorblockA{\textit{Department of Mechanical and Automation Engineering} \\
		\textit{The Chinese University of Hong Kong}\\
		Hong Kong SAR, China \\
		dkhan@mae.cuhk.edu.hk \\
		{*}Corresponding author}
		\and
\IEEEauthorblockN{2\textsuperscript{nd} Shihan Huang}
\IEEEauthorblockA{\textit{Department of Mechanical and Automation Engineering} \\
	\textit{The Chinese University of Hong Kong}\\
	Hong Kong SAR, China \\
	shhuang@link.cuhk.edu.hk}
		\and
\IEEEauthorblockN{3\textsuperscript{rd} Pak Kei Abia Hui}
\IEEEauthorblockA{\textit{Department of Mechanical and Automation Engineering} \\
	\textit{The Chinese University of Hong Kong}\\
	Hong Kong SAR, China \\
	abiahui@link.cuhk.edu.hk}
		\and
		\IEEEauthorblockN{4\textsuperscript{th} Yue Chen{*}}
		\IEEEauthorblockA{\textit{Department of Mechanical and Automation Engineering} \\
			\textit{The Chinese University of Hong Kong}\\
			yuechen@mae.cuhk.edu.hk \\
			{*}Corresponding author}
	}
	
	\maketitle

\begin{abstract}
Innovation and development of renewable energy devices are crucial for reaching a sustainable and environmentally conscious future. This work focuses on the development of a new type of renewable energy devices in the context of Smart Garden at the Chinese University of Hong Kong, which aims to design a bladeless wind turbine for urban areas, addressing the pressing need for clean energy locally and globally. Traditional wind turbines have been widely adopted in recent decades, while bladeless wind turbines have also displayed their advantages and uniqueness in urban areas. A Vortex Bladeless Wind Turbine (VBWT) is modeled by using Fusion 360 to optimize wind energy generation in urban settings with limited space and buildings-dominated landscape. Optimal parameters of the VBWT were obtained by comparing the results of drag force, lift force and deflection, via the simulations in Ansys. Hardware of proposed bladeless wind turbine has been assembled and developed by 3-dimensional printing. Additional tests and adjustments on hardware further improve the performance of the developed wind turbine. The outcomes of this work have the potential to contribute to future renewable energy initiatives and devote the sustainability efforts in urban energy systems.
\end{abstract}


\section{Introduction}
\IEEEPARstart{I}{ntroducing} urban wind energy provides a viable solution to the limited land availability and abundance of high-rise buildings \cite{reja22eb}. Traditional wind turbines are ineffective in turbulent urban wind conditions and low wind speeds in metropolitan areas. This work aims to develop a bladeless wind turbine specifically designed for urban areas, addressing the limited land situation and contributing to the development of renewable energy devices.

Wind turbines are devices that convert wind into electrical energy by harnessing its kinetic energy \cite{hart20wes}. Traditional wind turbines, such as horizontal-axis and vertical-axis turbines, have long been recognized as significant contributors to renewable energy. These turbines generate electricity by rotating their blades using wind energy, which can be utilized or stored in grid-connected systems. However, their large size and scale requirements make them expensive to install and maintain, posing a barrier to their use in urban areas. On the other hand, bladeless wind turbines offer a novel and innovative approach to capturing wind energy. Instead of using rotor blades, these turbines employ alternative mechanisms to capture wind energy through oscillation. By utilizing systems like alternators and turning mechanisms, which include permanent magnets and copper coils, they generate power from wind vibrations \cite{badri23e3s}. There are three main types of wind turbines: Horizontal-axis (HAWTs), vertical-axis (VAWTs), and bladeless turbines. Their comparison is given in Table \ref{table1}.

\begin{table*}\centering
	\caption{Comparison of HAWTs, VAWTs, and Bladeless wind turbines}
	\resizebox{\textwidth}{!}{
		\begin{tabular}{llll}
			\hline \toprule[1.1pt]
			\centering{Feature} & HAWTs & VAMTs & Bladeless Wind Turines \\ 
			\hline \toprule[1.1pt]
			Advantage & \begin{tabular}[c]{@{}l@{}} \tabitem Higher efficiency \\ \tabitem Wider range of usage \\ \tabitem Higher power output \end{tabular} & \begin{tabular}[c]{@{}l@{}} \tabitem Omnidirectional capability \\ \tabitem Compact design \end{tabular} & \begin{tabular}[c]{@{}l@{}} \tabitem Reduced noise and visual impact \\ \tabitem Easy maintenance \end{tabular} \\ \hline
			
			Disadvantage & \begin{tabular}[c]{@{}l@{}} \tabitem More visual and noise impact \\ \tabitem Wind direction dependency \\ \tabitem Large-scale requirement \end{tabular} & \begin{tabular}[c]{@{}l@{}} \tabitem Lower efficiency \\ \tabitem Limited power output \end{tabular} & \begin{tabular}[c]{@{}l@{}} \tabitem Lower efficiency \\ \tabitem Limited real-time data \end{tabular} \\
			\hline
			\toprule[1.1pt]
	\end{tabular}}
	\label{table1}
\end{table*}
Various types of bladeless wind turbines have been developed and implemented, including the Dyson bladeless wind turbine, Aeromine's wind turbine, and the Vortex Bladeless Wind Turbine (VBWT) \cite{tandel21jpcs}. The VBWT has emerged as an innovative and promising alternative to traditional wind turbines. Its unique design harnesses vortex shedding, offering advantages such as reduced maintenance and operational costs due to the absence of rotating blades. Research studies have investigated the feasibility and performance of VBWTs. Experimental tests on small-scale prototypes have demonstrated their ability to generate electricity under controlled wind conditions \cite{Sayed23energies}. Computational fluid dynamics simulations have provided insights into the turbine's aerodynamic characteristics and highlighted opportunities for design optimization \cite{Elsayed22we}. Moreover, the integration of VBWTs in urban environments has shown their suitability for decentralized energy generation due to their compact size and low noise levels \cite{francis21mt} . For the aerodynamic mechanism can be found in \cite{reba66sa,behroozi19tpt,paul22jw}.

\section{Main Results}

To achieve sustainable development and fulfill carbon neutrality in 2050, the VBWT is the proper solution for Greater Bay Area which is characterized by limited land and resources due to its small scale and fewer environmental impacts.

\subsection{Site and motor selection}

\begin{figure}[!t]
	\centering
	\begin{tikzpicture}[font=\scriptsize]  
		
		\begin{axis}  
			[  
			ybar,  
			enlargelimits=0.07,  
			ylabel={Wind speed [m/s]}, 
			xlabel={\ Month},  
			symbolic x coords={Jan., Feb., Mar., Apr., May, Jun., Jul., Aug., Sep., Oct., Nov., Dec., Ave.}, 
			xtick=data,  
			nodes near coords, 
			nodes near coords align={vertical},  
			xticklabel style={rotate=45, anchor=north east},
			]  
			\addplot coordinates {(Jan.,5.97) (Feb.,7.49) (Mar.,7.38) (Apr.,6.96) (May,6.53)(Jun.,9.85) (Jul.,9.64) (Aug.,6.08) (Sep.,6.03) (Oct.,8.38) (Nov., 6.55) (Dec.,7.4) (Ave.,7.35) };  
			
		\end{axis}
	\end{tikzpicture}  
	\caption{Average monthly wind speed in Sha Tin, Hong Kong SAR, China, in 2022.}
	\label{wind speed}
\end{figure}
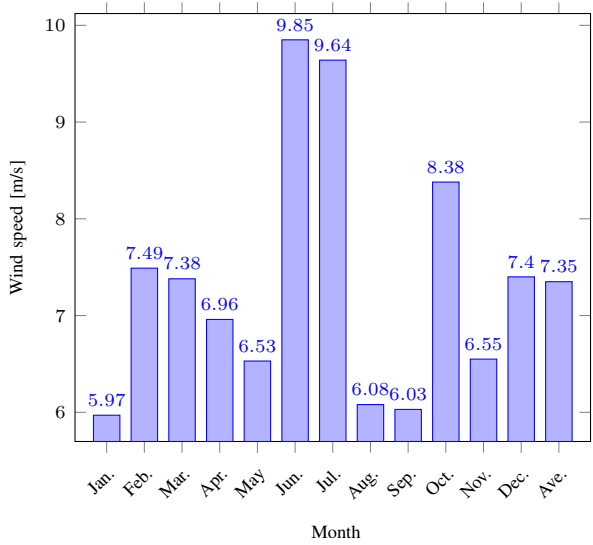

In the context of harnessing urban wind energy, it is momentous to consider the wind speed requirements of bladeless wind turbines. The optimal wind speed range for bladeless wind turbines is typically between 3 to 8 m/s. Similarly, VBWTs can effectively operate within a range of approximately 3 to 6 m/s. The average wind speed is relatively low in Hong Kong SAR, China. From Fig. \ref{wind speed}, the average wind speed in Sha Tin in 2022 is around 7.35 m/s. 

Buildings in urban areas create an urban wind acceleration effect, which occurs when the wind is channeled between buildings, resulting in increased wind speeds. Rooftop installations can take advantage of this accelerated wind, enhancing the energy generation potential of the turbine. Therefore, the above types of locations allow for the consistent and undisturbed flow of wind, resulting in optimal turbine performance and energy generation. Summarizing the above considerations, rooftops with low wind speed can be selected as appropriate places for installing bladeless wind turbines.

Brushed motor and brushless motor are two types of electric motor. The usage, efficiency, and lifespan are the main difference between brushed motors and brushless motors: Efficiency of brushed motor is relatively lower than brushless motor, while the maintenance of brushed motor is relatively higher than brushless motor. At the same time, the lifespan of brushed motor is comparatively shorter than brushless motor. In the comparison between brushed motors and brushless motors, the brushless (BLDC) motor is chosen for its superior efficiency and extended operational lifespan. Table \ref{table2} presents a detailed overview of the variances observed among BLDC motors with different numbers of poles.

\begin{table}\centering
	\caption{Comparison of different poles of BLDC motor}
		\begin{tabular}{llll}
			\hline \toprule[1.1pt]
			\centering{Motor type} & 3-pole motor & 6-pole motor & 9-pole motor \\ 
			\hline \toprule[1.1pt]
			Torque output & Moderate & High & Higher \\ \hline
			
			Speed range & High & Moderate & Lower \\ \hline
			
			Efficiency & Moderate & Moderate & High at low speeds \\ \hline
			
			Size & Compact & Moderate & Large \\ \hline
			
			Weight & Light & Moderate &  heavy \\ \hline
			
			Start-up speed & Moderate & Moderate & Low \\ \hline
			
			Power generation & Moderate & High & Higher \\ \hline
			
			Cost & Low & Moderate & High \\ \hline
			
			\toprule[1.1pt]
		\end{tabular}
	\label{table2}
\end{table}

In the context of bladeless wind turbine, 6 and 9-pole BLDC motors exhibit superior power generation and efficiency at low speeds, which aligns well with the requirements of such turbines. Conversely, 3 and 6-pole BLDC motors offer the advantages of lower cost and reduced weight. Consequently, when considering the specific case of a bladeless wind turbine, the 6-pole BLDC motor emerges as a potentially optimal choice.

\subsection{Key parameters selection}
Simulation is done by Ansys 2024 R1 version. Different parameters have been set for comparing the deflection, drag force and lift force.

A width of 1.5 mm was initially chosen for the bladeless wind turbine design. However, due to the weight of the turbine, it was decided that a narrower width of 1 mm would be more advantageous. This adjustment in width aims to achieve an optimal balance between the turbine's structural integrity and weight considerations. The selection of crucial parameters for the optimal turbine design follows a prioritized approach. The average deflection parameter is given highest priority as it directly impacts the turbine's structural integrity and stability. Minimizing deflection under wind load improves the turbine's overall performance and longevity. The lift force parameter is of secondary importance as it affects the turbine's efficiency and its ability to generate power by providing the necessary upward force. Lastly, the drag force parameter, though considered, is given lower priority compared to the other two parameters. Prioritization enables the identification and selection of the most favorable design, achieving a balance between structural integrity, power generation efficiency, and aerodynamic performance.

\begin{figure}
	\centering
	\begin{subfigure}{.25\textwidth}
		\centering
		\includegraphics[width=.9\linewidth]{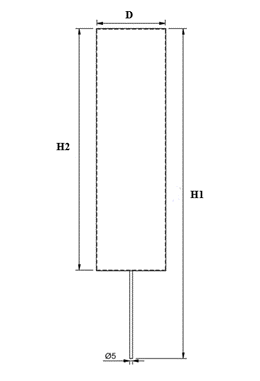}
		\caption{2D design sketch}
	\end{subfigure}%
	\begin{subfigure}{.25\textwidth}
		\centering
		\includegraphics[width=.6\linewidth]{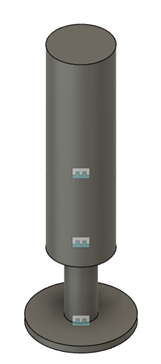}
		\caption{3D design sketch}
	\end{subfigure}
	\caption{Bladeless wind turbine design}
	\label{bwtd}
\end{figure}
The control group, referred to as Case 1, serves as the baseline for comparison, while experimental groups (Case 2 to Case 11) are created by setting specific parameters within a fixed range. In this simulation, the diameter ($D$), total height ($H_1$), and mast height ($H_2$) were analyzed in 50 mm intervals. However, due to the turbine's large size, conducting simulations using Ansys software was not feasible, preventing the testing of the maximum and minimum deflection of Case 5. To resolve this issue, the diameter parameter was adjusted to be tested in 25 mm intervals, ensuring successful acquisition of all experimental results. This modification allowed for a comprehensive evaluation of the desired parameters while accommodating the limitations of the simulation software.

\begin{table*}\centering
	\caption{Simulation results using Ansys 2024}
	\resizebox{\textwidth}{!}{
		\begin{tabular}{cccccccccccc}
			\hline \toprule[1.1pt]
			\centering{Mode} & Original Design & \multicolumn{4}{c}{Changed values of $D$} & \multicolumn{3}{c}{Changed values of $H_1$} & \multicolumn{3}{c}{Changed values of $H_2$}\\ 
			\hline
			Case number & 1 & 2 & 3 & 4 & 5 & 6 & 7 & 8 & 9 & 10 & 11 \\
			\hline \toprule[1.1pt]
			$D$ (mm) & 125 & 100 & 150 & 175 & 200 & 125 &	125 & 125 & 125 & 125 & 125\\
			\hline
			$H_1$ (mm) & 600 & 600 & 600 & 600 & 600 & 550 & 650 & 700 & 600 & 600 & 600\\
			\hline
			$H_2$ (mm) & 440 & 440 & 440 & 440 & 440 & 440 & 440 & 440 & 400 & 500 & 550\\
			\hline
			Max. Deflection (mm) & 41.69&34.25&46.94&56.30&N/A&26.26&57.98&85.40&46.00&33.57&21.73\\
			\hline
			Min. Deflection (mm)& 0&0&0&0&0&0&0&0&0&0&0\\
			\hline
			Average Deflection (mm)&23.62&19.28&26.47&31.92&N/A&14.32&33.97&51.58&27.18&17.82&11.24\\
			\hline
			Drag Force&0.610&0.506&0.700&0.816&0.917&0.628&0.593&0.613&0.564&0.731&0.800\\
			\hline
			Lift Force&0.009&0.006&-0.018&0.047&0.025&0.040&0.013&-0.023&0.026&-0.016&-0.025\\
			\hline
			Average rank&5&4&8&3&10&7&2&5&1&9&11\\
			\hline
			\toprule[1.1pt]
		\end{tabular}
	}
	\label{table3}
\end{table*}

\begin{figure}[ht]
	\centering
	\begin{subfigure}{.25\textwidth}
		\centering
		\includegraphics[width=.9\linewidth]{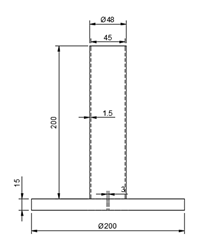}
		\caption{2D design sketch of base}
	\end{subfigure}%
	\begin{subfigure}{.25\textwidth}
		\centering
		\includegraphics[width=.6\linewidth]{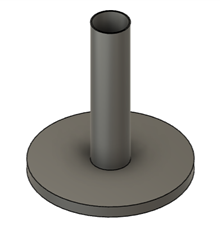}
		\caption{3D design sketch of base}
	\end{subfigure}
	\caption{Finalized base design}
	\label{fig:base}
\end{figure}

The simulation results provide clear conclusions regarding average deflection, drag force, and lift force. In terms of average deflection, a thorough analysis was conducted on various factors. Firstly, it was observed that variations in the diameter ($D$) had a noticeable impact on average deflection, with larger diameters resulting in higher average deflection values. Similarly, higher values of total height ($H_1$) were associated with increased average deflection. Additionally, investigations into different mast heights ($H_2$) revealed that smaller H2 values were correlated with higher average deflection (shown in Fig. \ref{bwtd}). These empirical findings highlight the importance of average deflection in influencing the structural behavior of bladeless wind turbines. Larger diameters, higher total heights, and smaller mast heights contribute to elevated levels of average deflection.

\begin{figure*}[ht]
	\centering
	\begin{subfigure}{.5\textwidth}
		\centering
		\includegraphics[width=0.9\linewidth]{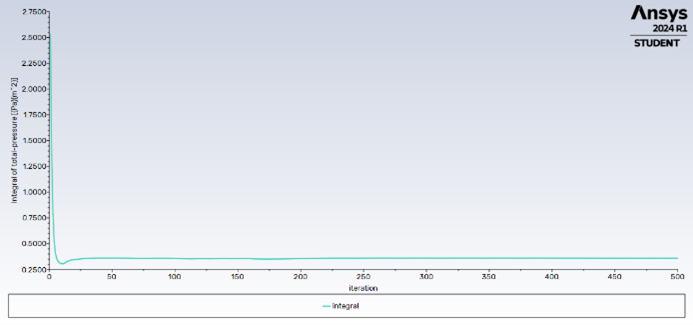}
		\caption{Integral}
	\end{subfigure}%
	\begin{subfigure}{.5\textwidth}
		\centering
		\includegraphics[width=0.9\linewidth]{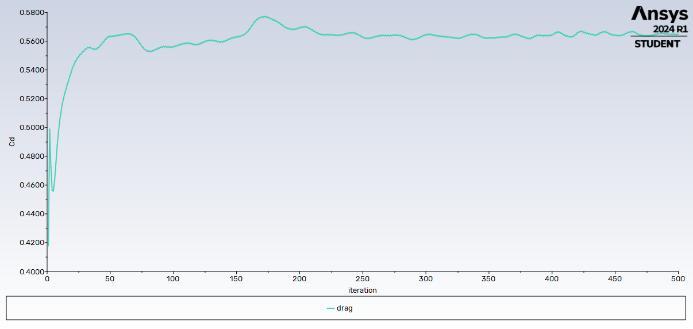}
		\caption{Drag}
	\end{subfigure}\\
	\begin{subfigure}{.5\textwidth}
		\centering
		\includegraphics[width=0.9\linewidth]{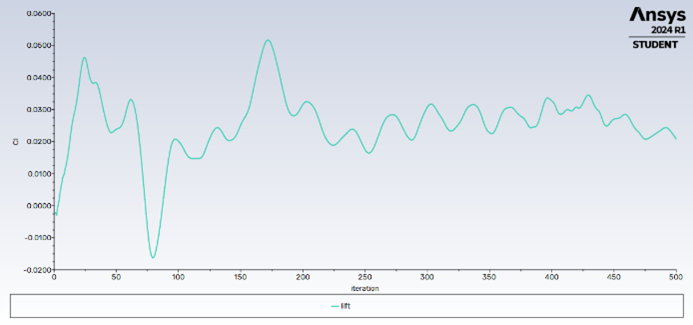}
		\caption{Lift}
	\end{subfigure}%
	\begin{subfigure}{.5\textwidth}
		\centering
		\includegraphics[width=0.9\linewidth]{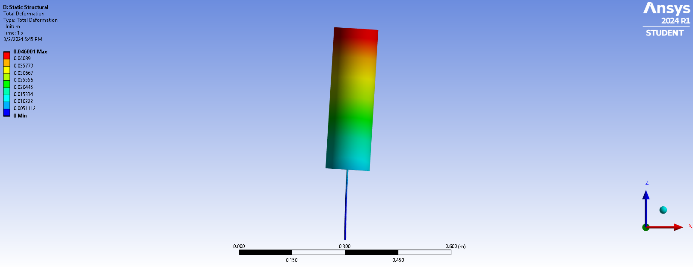}
		\caption{Deformation}
	\end{subfigure}
	\caption{Key parameters in the simulation of Case 9}
	\label{kp}
\end{figure*}
\begin{figure}[ht]
	\centering
	\begin{subfigure}{.25\textwidth}
		\centering
		\includegraphics[width=.9\linewidth]{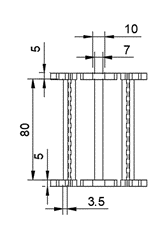}
		\caption{2D design sketch}
	\end{subfigure}%
	\begin{subfigure}{.25\textwidth}
		\centering
		\includegraphics[width=.6\linewidth]{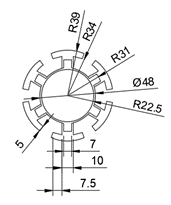}
		\caption{3D design sketch}
	\end{subfigure}\\
	\begin{subfigure}{.25\textwidth}
		\centering
		\includegraphics[width=.6\linewidth]{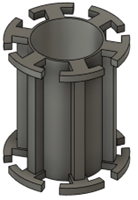}
		\caption{3D design sketch}
	\end{subfigure}
	\caption{Finalized design of coil case}
	\label{figure:cc}
\end{figure}

Considering the simulation results shown in Table \ref{table3}, Case 9 is identified as the most suitable design among all the simulation cases, taking into account average deflection, drag force, and lift force. Regarding the base size, improvements have been made to certain parts of the base, as shown in Fig. \ref{fig:base}. There are no specific material requirements for constructing the base and copper coil case (shown in Fig. \ref{figure:cc}), making white resin the optimal choice due to its cost-effectiveness and material hardness. The key parameters for the base are as follows: total height of 215 mm, upper height of 200 mm, bottom height of 15 mm, upper diameter of 48 mm, lower diameter of 200 mm, width of 1 mm, and a diameter of 3 mm for placing the rod, where results are shown in Fig. \ref{kp}.

\begin{figure}[ht]
	\centering
	\begin{subfigure}{.25\textwidth}
		\centering
		\includegraphics[width=.5\linewidth]{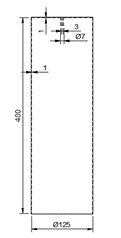}
		\caption{2D side view of mast}
	\end{subfigure}%
	\begin{subfigure}{.25\textwidth}
		\centering
		\includegraphics[width=.6\linewidth]{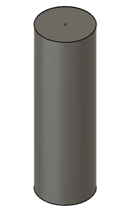}
		\caption{3D design of mast}
	\end{subfigure}
	\caption{Finalized design of mast}
	\label{fig:mast}
\end{figure}

Carbon fiber reinforced polymer has been employed in the construction of the mast to achieve a lightweight structure. Carbon fibers, known for their elevated tensile strength and stiffness, surpass the mechanical properties of glass fibers. Key parameters are selected as follows: Height: 440 mm, Diameter: 125 mm, Width: 1 mm, Diameter for placing rod: 3 mm. The finalized design of mast is shown in Fig. \ref{fig:mast}.

\subsection{Hardware testing}
\begin{figure}[ht]
	\centering
	\includegraphics[width=3.3in]{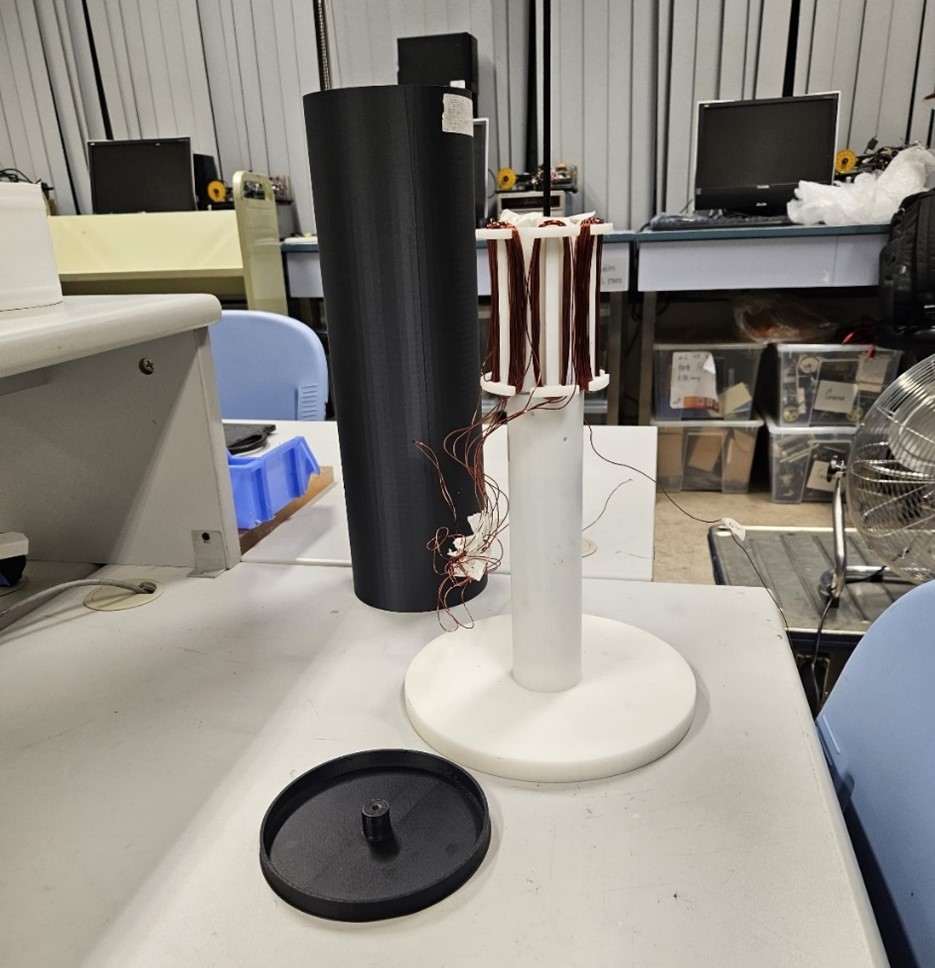}
	\caption{Printed components of bladeless wind turbine.}
	\label{fig:printed}
\end{figure}

The design in the previous subsection was sent to a 3D printing company to print the models. The printed mast, base and copper coil case, and the assembled bladeless wind turbine is shown in Fig. \ref{fig:printed}.

\begin{figure}[ht]
	\centering
	\includegraphics[width=3.3in]{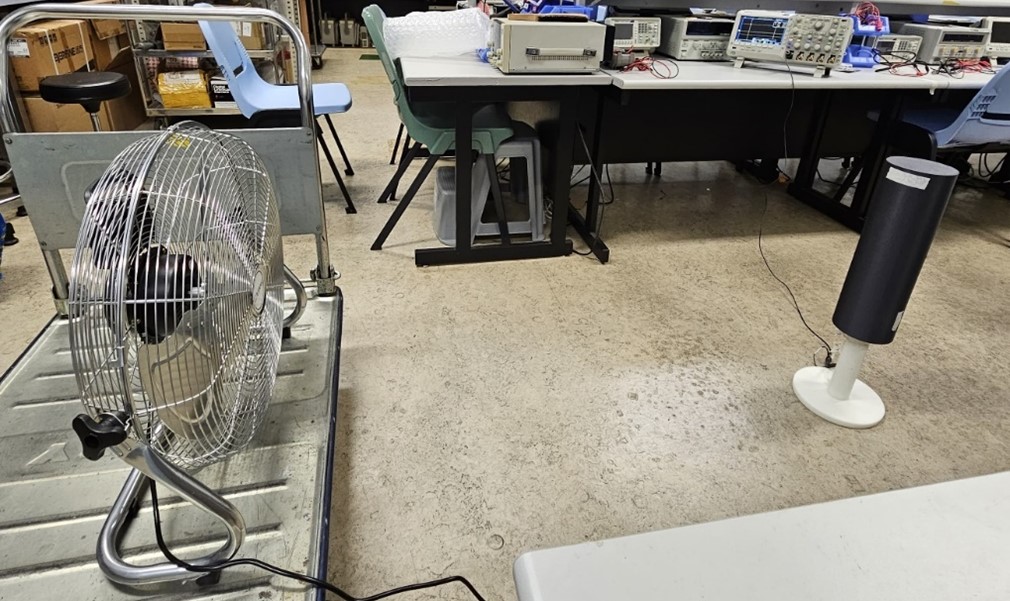}
	\caption{Testing environment set up.}
	\label{fig:testenv}
\end{figure}

Upon receiving and assembling the printed models, a preliminary testing phase was conducted to assess the bladeless wind turbine's optimal effectiveness and efficiency. This testing process aims to evaluate and verify the turbine's highest performance capabilities. The testing configuration for the experiment is illustrated in Fig. \ref{fig:testenv}. The energy generator consisted of 8 neodymium magnets and was accompanied by 6 sets of copper coils, each set containing 30 coils. Moreover, the distance between the bladeless wind turbine and the fan was approximately 1 meter. The testing process encompassed three distinct wind speeds: 4 m/s, 4.5 m/s, and 5 m/s. These varying wind speeds were crucial for evaluating the performance and efficiency of the bladeless wind turbine under different wind conditions. This comprehensive testing setup allowed for a thorough analysis of the turbine's capabilities and its ability to harness wind energy effectively.

\begin{figure}[ht]
	\centering
	\begin{subfigure}{.25\textwidth}
		\centering
		\includegraphics[width=.9\linewidth]{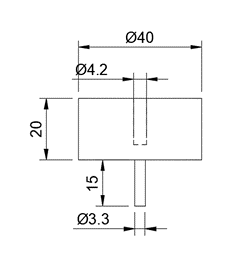}
		\caption{2D design sketch}
	\end{subfigure}%
	\begin{subfigure}{.25\textwidth}
		\centering
		\includegraphics[width=.6\linewidth]{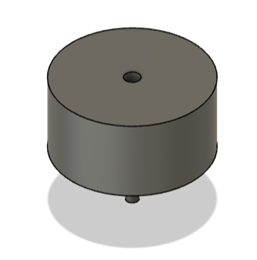}
		\caption{3D design sketch}
	\end{subfigure}\\
	\begin{subfigure}{.5\textwidth}
		\centering
		\includegraphics[width=.6\linewidth]{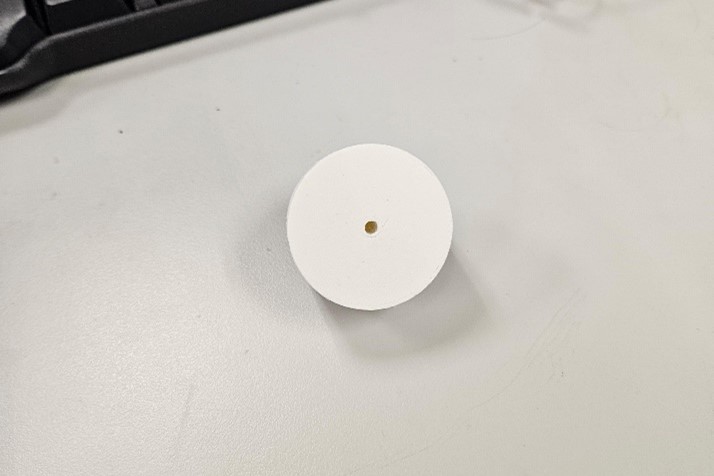}
		\caption{Printed rod in a modified version}
	\end{subfigure}
	\caption{Base for placing rod}
	\label{fig:pr}
\end{figure}

Since only one size of rod could be tested initially, a base was developed to enable the evaluation of vibration effectiveness using rods of different sizes. Through testing, it was determined that the 4 mm diameter rod exhibited the most favorable vibration performance among all tested sizes. As a result, the decision was made to use a 4 mm diameter rod for subsequent testing. This selection was based on the observation that the 4 mm diameter rod demonstrated superior vibrational characteristics compared to other sizes tested (shown in Fig. \ref{fig:pr}). This information will guide future testing efforts and allow for a comprehensive assessment of the rod's performance.

To further increase the power output, it is imperative to enhance the efficiency of the bladeless wind turbine. The original design incorporated 6 sets of copper coils, with each set comprising merely 30 coils. However, this coil configuration proved insufficient to generate significant energy. Consequently, an improved approach is proposed, involving a substantial increase in the number of coils to 3000 coils per set. This adjustment will provide a more robust method for evaluating the efficiency of the bladeless wind turbine. Furthermore, to prevent the copper coil from becoming excessively thick, a reduced diameter of 0.1 mm will be implemented by utilizing copper wire during subsequent testing. This modification aims to optimize the performance of the turbine while maintaining the desired physical characteristics of the copper coil.


Based on the improvements, it is suggested that the efficiency of the bladeless wind turbine can be enhanced by increasing the number of coils in each set to a total of 3000 coils. This adjustment aims to improve the accuracy of evaluating the turbine's efficiency. Additionally, based on the improvement observed during our test, the diameter of the rod would be modified to 4 mm. This alteration was anticipated to contribute to the overall enhancement of the turbine's performance. To ensure consistency in the testing process, the other parameters for assessing the efficiency of the bladeless wind turbine, such as distance and wind speeds, will remain identical to those employed in our test. By implementing these proposed modifications and maintaining consistency in the testing parameters, a more comprehensive evaluation of the turbine's efficiency can be achieved.

\section{Future works}
Moving forward, the future work on Vortex Bladeless Wind Turbines (VBWTs) focuses on several key areas to maximize their potential and address current limitations. Firstly, research efforts could be directed towards improving the turbine's efficiency and power generation capabilities. This could involve exploring different materials, refining the design, and optimizing the oscillation frequency to enhance energy conversion. Additionally, investigating the effects of various environmental conditions, such as wind speed and direction, on the turbine's performance will be crucial for its widespread adoption. Furthermore, studies should be conducted to evaluate the turbine's reliability and durability over extended periods, considering factors such as fatigue, wear, and maintenance requirements. Moreover, integrating energy storage systems with VBWTs could provide a solution for intermittent energy production, ensuring a more consistent and reliable power supply. Lastly, conducting comprehensive life-cycle assessments to assess the environmental impact, including carbon footprint and material waste, would aid in determining the turbine's sustainability and eco-friendliness. By addressing these future research areas, VBWTs can become a more viable and efficient renewable energy solution for a sustainable future.

\section{Discussion}
Admitting there exists room for the improvement of current model of bladeless wind turbines, it is worthwhile to explore potential ways for enhancing their performance. One promising approach is to increase the number of bladeless wind turbine installations, which can notably boost the overall efficiency of the system and lead to significant advancements in power generation capabilities. By scaling up the quantity of installed turbines, a larger amount of wind energy can be effectively harnessed and converted into clean electricity. This approach enables a more efficient utilization of wind resources as multiple turbines can capture wind from various directions and angles. Consequently, a comprehensive and consistent energy generation can be achieved, even under varying wind conditions. Moreover, a larger fleet of bladeless wind turbines facilitates the distribution of energy production across a wider area, potentially reducing transmission losses and enhancing overall grid stability. Therefore, by increasing the number of VBWTs installations, the efficiency of the entire system can be significantly improved, maximizing the utilization of wind resources and optimizing energy production. Further efforts could also be devoted to the formal analysis of coordination of multiple wind turbines in the context of control barrier functions and Lyapunov stability theory \cite{han19tac,han16gm}.

\section{Conclusion}
A new type of Vortex Bladeless Wind Turbine (VBWT) is designed to harness wind energy efficiently in small scale and low wind speed areas. By utilizing Fusion 360 and Ansys 2024 R1, the model of the bladeless wind turbine can be designed and simulated to identify optimal solutions and suitable parameters. Successful implementation of the bladeless wind turbine has demonstrated its ability to efficiently generate electrical power. However, it is admitted that there is space for improvement in terms of its efficiency. As a relatively new technology in the field of renewable energy, there is a limited amount of information, essays, and research available on bladeless wind turbines. Therefore, future research is essential to further enhance the development of bladeless wind turbines and other renewable energy devices. Future research will contribute to the advancement of bladeless wind turbine technology, facilitating its wider adoption and contributing to the future urban energy systems.

\section*{Acknowledgments}
Financial support from CUHK Research Committee Direct Grant for Research (4055228) and the Teaching Development and Language Enhancement Grant (4170989), Hong Kong SAR, are gratefully acknowledged. The authors appreciate valuable inputs from Prof. Prof. Pui Kuen Amos Tai and Dr. Kenneth Ming Li at The Chinese University of Hong Kong.

\vfill
\end{document}